ORIGINAL ARTICLE

# Detection of Fibrosis in Cine Magnetic Resonance Images Using Artificial Intelligence Techniques

*Detección de fibrosis en imágenes de cine de resonancia cardíaca mediante el uso de técnicas de inteligencia artificial*

ARIEL. H. CURIALE[1,2,3], FACUNDO CABRERA[2,3], PABLO JIMENEZ[2,3], JORGELINA MEDUS[4,5], GERMÁN MATO[2,3,6], MATÍAS E. CALANDRELLI[MTSAC, 2, 4]

**ABSTRACT**

**Background:** Artificial intelligence techniques have demonstrated great potential in cardiology, especially to detect imperceptible patterns for the human eye. In this sense, these techniques seem to be adequate to identify patterns in the myocardial texture which could lead to characterize and quantify fibrosis.
**Purpose:** The aim of this study was to postulate a new artificial intelligence method to identify fibrosis in cine cardiac magnetic resonance (CMR) imaging.
**Methods:** A retrospective observational study was carried out in a population of 75 subjects from a clinical center of San Carlos de Bariloche. The proposed method analyzes the myocardial texture in cine CMR images using a convolutional neural network to determine local myocardial tissue damage.
**Results:** An accuracy of 89% for quantifying local tissue damage was observed for the validation data set and 70% for the test set. In addition, the qualitative analysis showed a high spatial correlation in lesion location.
**Conclusions:** The postulated method enables to spatially identify fibrosis using only the information from cine nuclear magnetic resonance studies, demonstrating the potential of this technique to quantify myocardial viability in the future or to study the lesions etiology

**RESUMEN**

**Introducción:** Las técnicas de inteligencia artificial han demostrado tener un gran potencial en el área de la cardiología, especialmente para identificar patrones imperceptibles para el ser humano. En este sentido, dichas técnicas parecen ser las adecuadas para identificar patrones en la textura del miocardio con el objetivo de identificar y cuantificar la fibrosis.
**Objetivos:** Proponer un nuevo método de inteligencia artificial para identificar fibrosis en imágenes cine de resonancia cardíaca.
**Materiales y métodos:** Se realizó un estudio retrospectivo observacional en 75 sujetos del Sanatorio San Carlos de Bariloche. El método propuesto analiza la textura del miocardio en las imágenes cine CMR (resonancia magnética cardíaca) mediante el uso de una red neuronal convolucional que determinar el daño local del tejido miocárdico.
**Resultados:** Se observó una precisión del 89% para cuantificar el daño tisular local en el conjunto de datos de validación y de un 70% para el conjunto de prueba. Además, el análisis cualitativo realizado muestra una alta correlación espacial en la localización de la lesión.
**Conclusiones:** El método propuesto permite identificar espacialmente la fibrosis únicamente utilizando la información de los estudios de cine de resonancia magnética nuclear, mostrando el potencial de la técnica propuesta para cuantificar la viabilidad miocárdica en un futuro o estudiar la etiología de las lesiones.

**Palabras clave:** Redes Neuronales – Viabilidad Miocárdica – cine CMR – Radiómica – Fibrosis

## INTRODUCTION

Artificial Intelligence (AI) techniques, especially those based on deep neural networks, have shown great potential in the area of cardiology, as described in a recent Journal of the American College of Cardiology publication by Dey et al. (1) This review lists the achievements of these techniques whose applications range from electrocardiogram (ECG) signal analysis to cardiac nuclear magnetic resonance (CMR) imaging to obtain volumetric quantification of cardiac func-tion, study myocardial viability, diagnose arrhythmias, infarctions or valvular diseases. (2-7) Deep learning techniques manage to identify complex patterns, sometimes imperceptible to a specialist, directly from the data. These characteristics, similar to radiomic features that are extracted by texture analysis of images, can be used for diagnosis and/or quantification of various diseases. (8,9)

The hypothesis of this work and of radiomics in general, is that properly identified image characteristics can be useful in predicting prognosis and therapeutic response for various conditions, thus providing valuable information for personalized therapy. (10)

Cine CMR imaging is considered the gold standard for quantifying volumes, mass, and function of both ventricles. However, additional sequences are required for tissue characterization to identify edema and fat, among others. In the case of fibrosis, specific techniques are required, such as late gadolinium enhancement (LGE) sequences, which need additional time and the use of gadolinium since it is not possible to identify it in cine images. The possibility of acquiring useful information imperceptible to the expert eye through deep learning and radiomic techniques is attractive, since sequences that require contrast could be dispensed with in the future. Consequently, this study aims to address this problem using an automatic learning or Machine Learning system based on convolutional neural networks (CNN) to identify relevant characteristics in the myocardial texture present in cine CMR images, and thus enable the identification of fibrosis.

### OBJECTIVES

The aim of this work was to postulate a new AI technique to identify fibrosis in cine CMR images.

### METHODS

A retrospective observational study was carried out in 75 subjects [52±17 years old, 52 (72%) male] from Sanatorio San Carlos de Bariloche (SSC), suffering from various diseases: acute myocardial infarction (AMI), left ventricular hypertrophy (LVH), hypertrophic cardiomyopathy (HCM), dilated left ventricle (LV), and those with a normal diagnosis. Among the 75 subjects, 35 had myocardial lesions. Cine CMR images were acquired over a 2-year period with a Philips Intera 1.5 T scanner using a gated SSFP sequence.

Gadolinium 0.1 mmol/kg was used as contrast agent to identify and delineate damaged myocardial tissue. Lesion demarcation was performed by a specialized cardiologist from SSC using the Segment software in the LGE studies and automatically transferred to the cine CMR studies, identifying the time-instant or the closest time of the cardiac phase corresponding to the LGE study. Using automatic alignment of the ventricular centers and appropriate resampling, the scars in the cine CMR studies were identified. Finally, an analysis of the spatial concordance of the lesions observed in LGE was performed by a specialist. Overall, 73 subjects were used as 2 studies were excluded due to misalignment. The resolution of cine CMR images is on average 256 x 256 pixels with a spacing of 1.25 mm, while LGE studies have a resolution of 512 x 512 pixels with a spacing of 0.625 mm. The resolution along the long axis of the LV is 10 mm for both studies (cine CMR and LGE); however, the cine CMR studies have 12 slices and those of LGE have 9.

#### Postulated method
The postulated method analyses the myocardial texture in cine CMR images using a convolutional neural network (CNN). From the cine imaging of a subject, we proceed to identify the left ventricular myocardial tissue using the approach proposed by our group. (3,11) Once the myocardium has been identified, it is subdivided into regions or patches of 11×11 pixels, and the distance from the center of the patch to the center of the left ventricular mass is recorded, as can be seen in Figure 1. Then, these regions are individually analyzed by a convolutional neural network (Encoder+MLP) to identify the probability of having a lesion. It is important to note that the patch position information is entered in the last dense layer of the multilayer perceptron (MLP), as visualized in Figure 1 (Patch position). Finally, information from all regions is combined for quantification and visualization of tissue damage in the myocardium, which is normalized to the range [0, 1].

#### Ethical considerations
The study was carried out in compliance with the National Law for the Protection of Personal Data No. 25 326, safeguarding the identity of the patients and all their personal data. All sensitive information was duly anonymized. The study was conducted in accordance with national and regional ethical regulations, and its protocol was approved by the Ethics Committee of the Ministry of Health of the Province of Río Negro.

### RESULTS
Two sets of experiments were performed to analyze the proposed model. The first aimed to determine the size of the myocardial region that the model would ultimately use, while the second focused on studying the local accuracy of the model in detecting the presence of a lesion. In turn, the data set consisting of 73 subjects was subdivided into 3 subsets (training, validation and test) to study the power of generalization of the proposed model. The subdivision was performed randomly, maintaining the original distribution of damaged tissue in each subset. In this way, the training set was made up of 53 subjects (72% of the total), while the validation and test sets consisted of 10 subjects (14%) each.

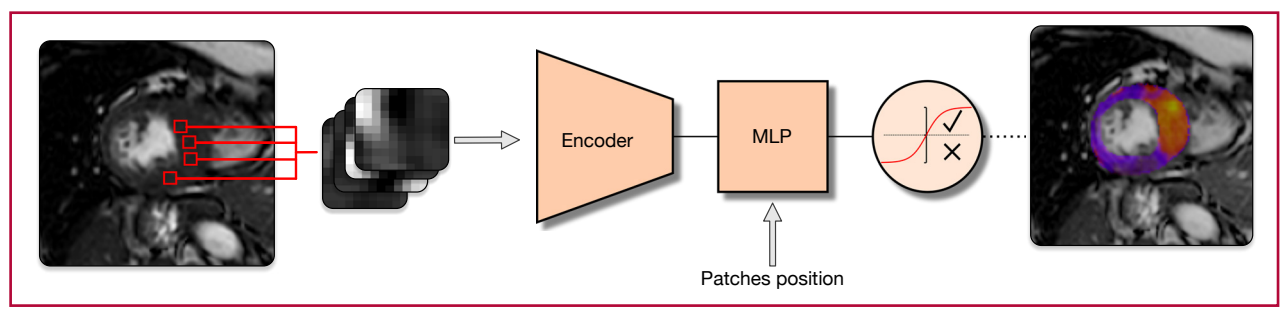

**Fig. 1**. Workflow used to quantify myocardial tissue damage in cine CMR images. See text for explanation (Methods section)

The results obtained when analyzing the size of the myocardial region (5×5, 7×7, 9×9 and 11×11 pixels), showed that a patch size of 11×11 pixels was the most adequate, obtaining the best result with an accuracy of 78% for the validation set. Importantly, sizes larger than 11×11 were excluded from the analysis because they significantly reduced the number of patches, and more significantly, the proportion of patches containing 50% of damaged tissue. The selection of the region of interest size at 11 pixels allowed the generation of a total of 2395 patches (50% with lesion and 50% without lesion), 1969 of which were used for training, 317 for validation and 109 to study the power of generalization of the proposed model. Next, the accuracy of the model was studied using data augmentation (patches' reflections and rotations) and the inclusion of the center of the patch with respect to the center of mass of the LV. The results show that the accuracy of the model is increased by 4% when using data augmentation, and an additional 5% when considering the information of the center of the patch with respect to the center of mass of the LV. Thus, the final model reaches an accuracy of 89% for the validation data and 70% for the test data. Figure 2 presents an example of tissue damage quantification for a subject with (a) and without (b) myocardial lesion. In turn, and as reference, the Figure shows cine CMR and LGE studies together with cine CMR plus manual delineation of the lesion superimposed on the cine CMR image. This result evidences a great spatial correlation of the zone that contains the lesion in the myocardium, exhibiting almost null quantification for the subject that does not present lesion.

### DISCUSSION

This work shows that it is possible to identify fibrosis in the myocardium from cine CMR images using AI techniques. Although a purely radiomic approach could have been chosen, this time we propose a model based on neural networks. The postulated model obtains an acceptable precision to identify fibrosis in cine sequences, which in clinical practice requires late enhancement sequences after gadolinium injection. The contribution of this study and the main advantage of the postulated method lie in the potential of this technique to identify whether a patient has myocardial fibrosis, without resorting to a late enhancement sequence. Thus, it is possible to reduce the time and cost of the study in those patients who do not have fibrosis. The proposed technique is not intended to replace LGE sequences to quantify fibrosis, but instead has the objective of avoiding carrying out such a study in patients who do not have fibrosis.

The idea of exploring tissue texture and using AI techniques to obtain a prediction is a concept that has been investigated from radiomics in oncology imaging, and has recently generated interest in the cardiovascular field, especially with cine CMR imaging. Although this study, which represents a proof of concept, proposes a technique that specifically identifies fibrosis in cine images, it is not difficult to imagine the potential of the proposed method, since in the future it could be extended to the identification of other tissue abnormalities such as edema, fat or increased extracellular volume.

Supervised machine learning techniques, such as the one postulated in this work, strongly depend on the quantity and extension of the data used for their training. In this sense, it is observed that it tends to overestimate the areas without tissue damage and underestimate those with significant deterioration, as can be seen in Figure 2. To improve the accuracy of the proposed method, it is necessary to extend the study to a larger number of cases. Finally, it is necessary to clarify that the fibrosis detected in this study does not include diffuse fibrosis, currently evaluated with T1 mapping sequences.

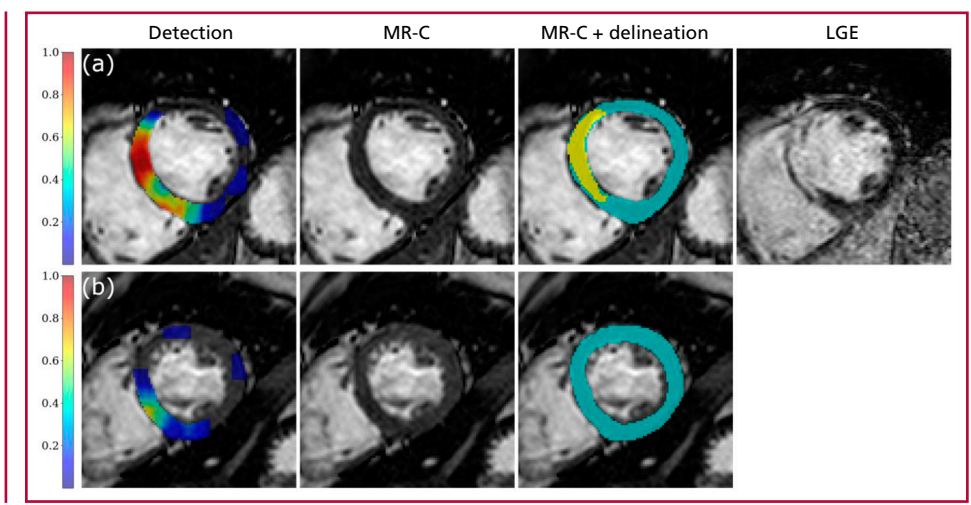

**Fig. 2.** Example of myocardial fibrosis detected in cine CMR images. Row (a) shows a patient with fibrosis as revealed by the LGE study (LGE column) and average detection greater than 0.5 (Detection column), while row (b) shows another subject without lesion whose average detection value is less than 0.5.

### Limitations

A decrease of almost 20% in precision was observed when using the test data (89% vs. 70%) which indicates the need to extend the study. As mentioned in the discussion, a study with a larger number of cases, ideally multicenter studies, where different devices are used and patients present myocardial lesions of different etiologies, is required to increase and validate the accuracy of the proposed model.

### CONCLUSIONS

To the best of our knowledge, this is the first study in the country that evaluated artificial intelligence techniques to quantify myocardial tissue lesions using only cine CMR imaging. In this work, it was possible to identify fibrosis in cine CMR images using AI techniques with acceptable precision. With the development of these techniques, the need for sequences that include gadolinium contrast to identify those patients with tissue damage could be avoided in the future. However, it is necessary to extend this work to reduce the observed error, improve the precision and validate the proposed methodology in a larger population.